\documentclass[12pt]{iopart}

\usepackage{iopams}
\usepackage{psfig}

\begin{document}

\jl{4}

\title[Observations of TeV gamma rays at large zenith
angles]{Observations of TeV gamma rays from Markarian 501 at large
zenith angles}

\author{P M Chadwick, K Lyons, T J L McComb\footnote[1]{To whom
correspondence should be addressed.}, K J Orford,\\ J L Osborne, S M
Rayner, S E Shaw and K E Turver}

\address{Department of Physics, Rochester Building, Science
Laboratories, University of Durham, Durham DH1 3LE, UK}

\begin{abstract} 

TeV gamma rays from the blazar Markarian 501 have been detected with the
University of Durham Mark 6 atmospheric \v{C}erenkov telescope using the
imaging technique at large zenith angles. Observations were made at
zenith angles in the range $70^{\circ} - 73^{\circ}$ during 1997 July
and August when Markarian 501 was undergoing a prolonged and strong
flare.

\end{abstract}

\pacs{95.85.Pw, 98.54.Cm, 98.62.Js, 98.70.Rz}

\submitted

\maketitle


\section{Introduction}

The use of atmospheric \v{C}erenkov telescopes to make observations at
large zenith angles was suggested by Sommers and Elbert
\cite{kn:sommers1987} as a potentially efficient method of measuring
gamma rays at the highest energy. The underlying principle is that at
large zenith angles, the footprint of an air shower as seen in optical
\v{C}erenkov light (which is a penetrating component of the cascade)
becomes large. This means that, although the threshold energy of a
telescope will increase as the zenith angle increases, the accompanying
increase in collecting area results in a useful gamma ray detection
rate. Sommers and Elbert discussed this large zenith angle technique in
the context of first-generation (non-imaging) gamma ray telescopes. 

Markarian 501 (Mrk 501) is a relatively close BL Lac object (at a red
shift $z$ of 0.034) which shows strong, variable, non-thermal emission
from radio to X-ray wavelengths. Since its discovery as a VHE gamma ray
source \cite{kn:quinn1996}, Mrk 501 has been extensively monitored in
the VHE waveband. In 1997, the object went into a prolonged active state
at all wavebands. The VHE activity was detected by a number of Northern
hemisphere facilities employing imaging techniques (e.g.
\cite{kn:catanese1997,kn:aharonian1997,kn:hayashida1998,kn:barrau1997}).
This state of high activity lasted from 1997 March -- October and the
source exhibited a complex time history with flaring activity on a time
scale as short as a few hours \cite{kn:aharonian1998a}. Multiwavelength
studies of Mrk 501 during this time have been reported
\cite{kn:catanese1997} and the energy spectrum at TeV energies has been
established \cite{kn:aharonian1998a,kn:samuelson1998,kn:krennrich1998}.

The upper bound to the TeV energy spectrum of an AGN is of considerable
interest as it, together with the distance to the source, leads to a
value for the density of intergalactic infra red photons (see e.g.
\cite{kn:stecker1992,kn:biller1998,kn:funk1998}). The highest energy
photons observed from this source to date have been detected by the
HEGRA group and are in excess of 25 TeV \cite{kn:aharonian1998b}.

We report here observations of Mrk 501 made from the Southern hemisphere
at zenith angles in excess of $70^{\circ}$ using the University of
Durham Mark 6 imaging telescope. Our measurements of cascades which
develop through three atmospheres demonstrate that imaging techniques
may be applied at these large zenith angles with the prospect of
improving the significance of the gamma ray signal at high energies.

\section{Observations}

\subsection{The Mark 6 Telescope}

The University of Durham Mark 6 telescope is located at Narrabri NSW
Australia and has been described in detail \cite{kn:armstrong1998}. It
uses the established imaging technique to separate gamma rays from
background cosmic rays and incorporates a robust, noise-free trigger
involving signals from three parabolic flux collectors of diameter 7 m
mounted on a single alt-azimuth platform. The imaging camera comprises
109 pixels (91 with $0.25^{\circ}$ diameter and 18 with $0.5^{\circ}$
diameter) and is mounted at the focus of the central flux detector. It
combines with two low resolution cameras (19 pixels each of diameter
$0.5^{\circ}$) which are mounted at the foci of the left and right flux
collectors to provide a trigger demanding a simultaneous temporal (10 ns
gate) and spatial ($0.5^{\circ}$) coincidence of the \v{C}erenkov light
detected in the three cameras. This trigger system allows the telescope
to detect small light flashes, and hence low energy gamma rays, free of
the triggers due to light produced by individual local muons. This is
particularly important for observations at very large zenith angles when
the background from local muons mimicking the signatures of gamma rays
can become important.

\subsection{Observations of Mrk 501}

Mrk 501 was observed in 1997 July and August under clear and moonless
skies during an interval of extreme activity when detection rates for
gamma rays $> 300$ GeV (using conventional imaging telescopes) exceeded
10 photons per min (see e.g. Bradbury \cite{kn:bradbury1997b}).

Data were recorded at zenith angles of $70^{\circ} - 73^{\circ}$ using
the standard Mark 6 telescope procedure which involves recording data in
15 min segments. OFF source observations were taken by alternately
observing regions of the sky which differ by $\pm 15$ min in right
ascension from the position of Mrk 501 so ensuring that the ON and OFF
segments possess identical azimuth and zenith profiles. This observing
pattern, which involves reversal of the order of the ON and OFF source
measurements, eliminates any first order changes in telescope
performance due to residual secular changes in atmospheric clarity,
temperature etc. An observing log is shown in table
\ref{table:observing_log}. 

The background count rate of the telescope at $70^{\circ}$ to the zenith
was about 70 cpm, some 10\% of that at the directions close to the
zenith where the telescope would normally be operated. A preliminary
estimate of the threshold for gamma ray detection with the Mark 6
telescope at $70^{\circ}$ to the zenith is $\sim 15$ TeV which may be
compared with the value of $\sim 300$ GeV for zenith angles $ \leq
30^\circ$. We note that Sommers and Elbert \cite{kn:sommers1987}
estimate a factor of $\sim \times 70$ in threshold energy difference
between zenith angles of $30^\circ$ and $70^\circ$ compared with the
value of $\sim \times 50$ from our estimate.

\begin{table}

\caption{Observing log for our observations of Markarian 501 during
1997.\label{table:observing_log}}
\begin{indented}
\item[]\begin{tabular}{@{}lc}
\br
Date & No. of scans\\ \ns
& ON source \\ 
\mr

1997 July 4 & 3\\
1997 July 5 & 5\\
1997 July 7 & 4\\
1997 July 9 & 3\\
1997 July 23 & 1\\
1997 July 28 & 5\\
1997 July 30 & 5\\
1997 July 31 & 3\\
1997 August 1 & 4\\
1997 August 5 & 4\\
\br
\end{tabular}
\end{indented}

\end{table}

Data were accepted for analysis only if the sky was clear and stable
according to the telescope background count rate and a boresighted FIR
radiometer which provides a measure of sky clarity
\cite{kn:buckley1998}. The gross counting rates in each pair of ON-OFF
segments used were consistent at the $2.5 \sigma$ level. A total of 9.25
hrs of data for ON-source observations and an equal amount OFF source
meet these criteria.

\section{Results}

The reduction and analysis of accepted data follows a well established
routine \cite{kn:chadwick99}. The gains and pedestals of all 147 PMTs
and digitizer electronics are calibrated within each 15 min segment
using embedded laser and false coincidence events. PMT noise is
equalised for the ON and OFF source segments using the software padding
technique \cite{kn:cawley1993} prior to identifying the accurate
location of the source in the camera's field of view using the axial CCD
camera. Events confined to within $1.1^{\circ}$ of the centre of the
camera and which have in excess of 600 digital counts (to ensure
reliable image reconstruction) are considered suitable for further
analysis. Finally, the spatial moments of each image relative to the
source position are evaluated and those events are rejected which are
unlikely to have been initiated by gamma rays.

In addition to the background rejection based upon the image analysis, a
measure of the fluctuations between the centroids of the samples
recorded by the left and right collectors of the Mark 6 telescope
($D_{\rm dist}$) provide an additional discriminant
\cite{kn:chadwick1998a}. Gamma rays are identified on the basis of image
shape and left/right fluctuation and then plotting the number of events
as a function of the pointing parameter {\it ALPHA}. Gamma ray events
from a point source appear as an excess at small values of {\it ALPHA}.

The selection criteria appropriate to observations at large zenith
angles ($> 70^{\circ}$) are summarised in table \ref{select_table}.
Small changes have been allowed from the standard set of parameters
identified for the analysis of data taken with the Mark 6 telescope at
zenith angles $< 45^{\circ}$ to accommodate e.g. the narrower images of
cascades propagated through three atmospheres.

\begin{table}

\caption{The image parameter selections applied to the Markarian 501
data recorded at zenith angles between $70^\circ$ and $73^\circ$.
\label{select_table}}

\begin{indented}
\item[]\begin{tabular}{@{}lccccc}

\br
Parameter&Ranges&Ranges&Ranges&Ranges\\
\mr
{\it SIZE} (d.c.)&$600-800$&$800-1200$&$1200-1500$&$1500-5000$\\
{\it DISTANCE}&$0.35^{\circ}-0.85^{\circ}$&$0.35^{\circ}-0.85^{\circ}$&$0.35^{\circ}-0.85^{\circ}$&$0.35^{\circ}-0.85^{\circ}$\\
{\it ECCENTRICITY}&$0.35-0.85$&$0.35-0.85$&$0.35-0.85$&$0.35-0.85$\\
{\it WIDTH}&$ < 0.14^{\circ}$&$ < 0.18^{\circ}$&$ < 0.22^{\circ}$&$ < 0.22^{\circ}$\\
{\it CONCENTRATION}&$ < 0.70$&$ < 0.70$&$ < 0.70$&$ < 0.30$\\
$D_{\rm dist}$&$ < 0.20^{\circ}$&$ < 0.14^{\circ}$&$ < 0.12^{\circ}$&$ < 0.10^{\circ}$\\
\br

\end{tabular}

\end{indented}

\end{table} 

\begin{figure}[tb]

\centerline{\psfig{file=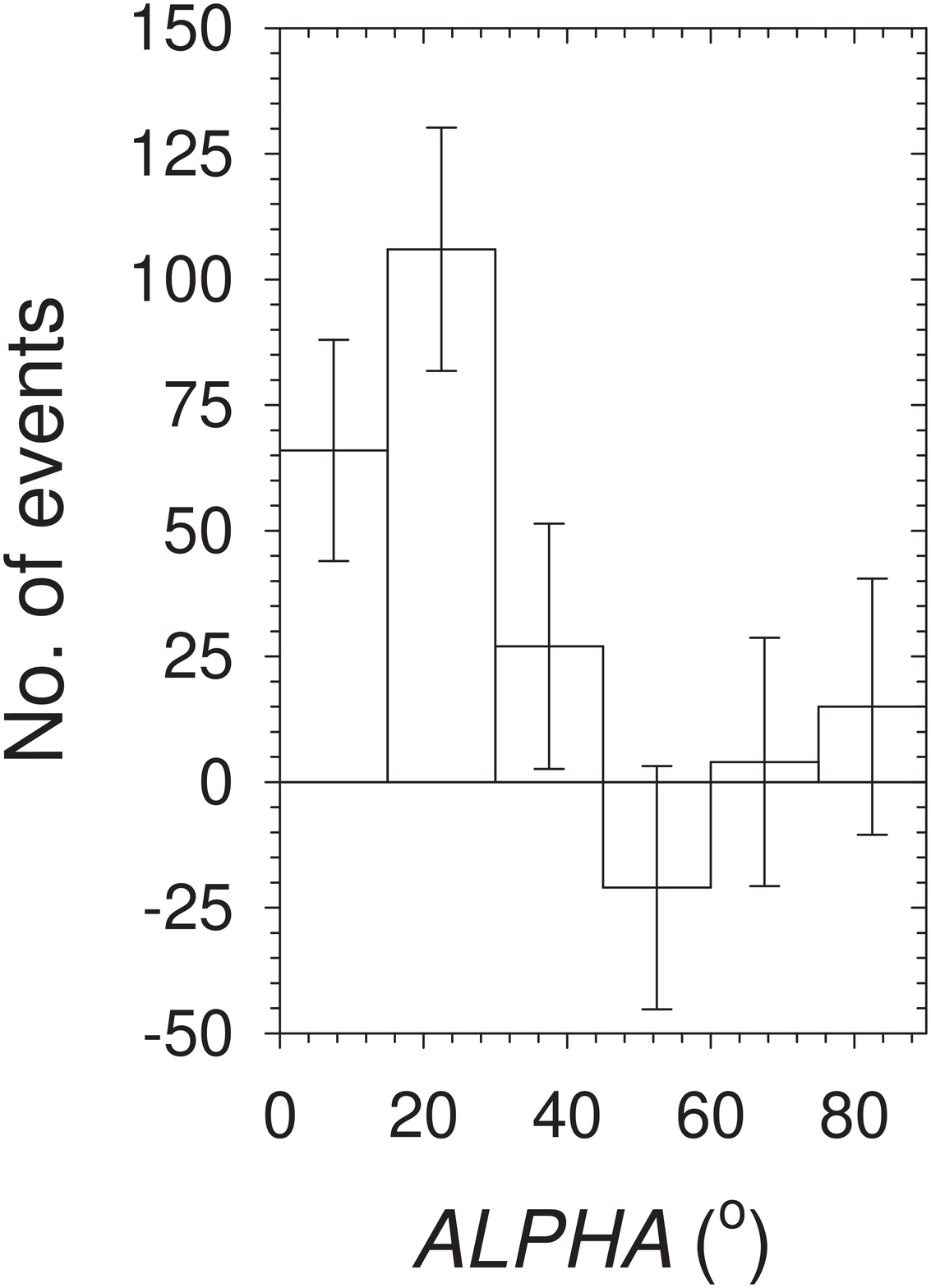,height=10cm}}


\caption{The difference in the distribution of values of {\it ALPHA} for
the shape selected data recorded ON and OFF source from Mrk 501.
\label{fig:alphaplot}}

\end{figure}

The number of events in the ON and OFF samples after the application of
the selections described above are summarised in table
\ref{result_table}. The {\it ALPHA} plot of the differences of the ON
and OFF distributions is shown in figure \ref{fig:alphaplot}. The excess
of events at small values of {\it ALPHA} suggest the detection of a
source with an excess of events with {\it ALPHA} $< 30^{\circ}$ at a
significance of $5.6 \sigma$. The width of the {\it ALPHA} distribution,
with excess events with {\it ALPHA} of up to $30^{\circ}$, may be a
consequence of the effects of magnetic field of $~ 0.4$ G, acting on
the cascade over unusually large linear distances \cite{kn:chadwick99a}.

\begin{table}

\caption{The results of various event selections for the Markarian 501
data. \label{result_table}}

\begin{indented}
\item[]\begin{tabular}{@{}lrrrr}

\br
& On & Off & Difference & Significance \\
\mr
Number of events & 40869 & 41232 & $-363$ & $-1.3~\sigma$ \\
\\
Number of size and & 21570 & 21286 & 284 & $1.4~\sigma$ \\ \ns
distance selected events & & & & \\
\\
Number of shape & 1935 & 1750 & 185 & $3.1~\sigma$ \\ \ns
selected events & & & & \\
\\
Number of shape and & 647 & 475 & 172 & $5.6~\sigma$ \\ \ns
{\it ALPHA} selected events & & & & \\
\br

\end{tabular}

\end{indented}

\end{table}

\begin{figure}[tb]

\centerline{\psfig{file=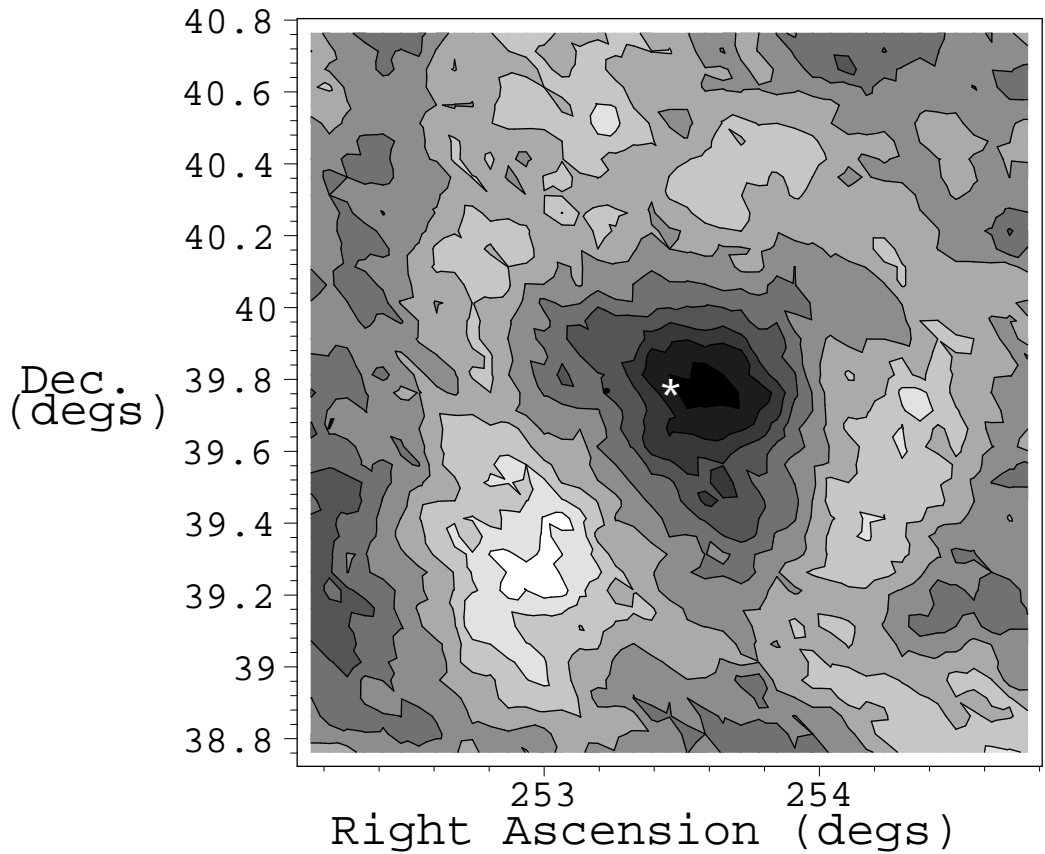,height=12cm}}

\caption{False source analysis: excess events with {\it ALPHA} $ <
30^\circ$. The contours are spaced at $1.0 \sigma$ intervals, with the
grey scale being such that black indicates a probability of $>
6\sigma$ for a gamma ray originating from that direction. The position
of Mrk 501 is marked by *.} \label{fig:mapleplot}

\end{figure}

The ON and OFF source data have been re-analysed using a matrix of
assumed source positions to provide a ``false source'' analysis. In
figure \ref{fig:mapleplot} we show that the excess of gamma ray events
originates from the source which is located in the right ascension,
declination plot at a position consistent with Mrk 501. (The camera
centre does not coincide with the source position.)

\section{Conclusion}

Observations of Mrk 501 during outburst made at zenith angles $>
70^\circ$ using a Southern hemisphere imaging telescope have
demonstrated the efficacy of the large zenith angle ACT technique
suggested by Sommers and Elbert \cite{kn:sommers1987}.

The selection criteria used to reject background in observations with
our imaging telescope at large zenith angles are similar to those
employed for observations at small zenith angles. The main difference is
in the maximum value of the width of images accepted, which are smaller
than those in observations near the zenith due to the increased distance
to cascade maximum.

In the absence of a network of gamma ray observatories giving full sky
coverage, this detection of Mrk 501 has demonstrated that observations
at large zenith angles may allow more continuous monitoring of strong
TeV gamma ray sources.

Pending completion of a simulation study of the detailed response of the
Mark 6 telescope at zenith angles of $\sim 70^\circ$ it is not possible
to ascribe a reliable value to the highest energy gamma rays detected in
this short exposure observation. Thus it has not yet been possible to
extend the spectrum beyond the limiting energy available from more
extensive observations using atmospheric \v{C}erenkov telescope
techniques at smaller zenith angles from Northern hemisphere
observatories.

\ack

We are grateful to the UK Particle Physics and Astronomy Research
Council for support of the project and the University of Sydney for the
lease of the Narrabri site. The Mark 6 telescope was designed and
constructed with the assistance of the staff of the Physics Department,
University of Durham. The efforts of Mrs S E Hilton and Mr K
Tindale are acknowledged with gratitude.

\end{document}